\documentclass[preprint2]{aastex}
\usepackage{graphicx}
\usepackage{color}

\slugcomment{To be submitted to \emph{Astrophysical Journal}}

\begin{document}

\title{ Ultra High Energy Electrons Powered by Pulsar Rotation }

\author{Mahajan S.}
\affil{Institute for Fusion Studies, The University of Texas at
Austin, Austin, Texas 78712}

\author{Machabeli G.}
\affil{Centre for Theoretical Astrophysics, ITP, Ilia State
University, 0162-Tbilisi, Georgia}

\author{Osmanov Z.}
\affil{Free University of Tbilisi, 0183-Tbilisi, Georgia}

\author{Chkheidze N.}
\affil{Centre for Theoretical Astrophysics, ITP, Ilia State
University, 0162-Tbilisi, Georgia}
%\altaffiltext{2}{Also at the Free University of Tbilisi}
%\altaffiltext{3}{author3@astro-ge.org}

\begin{abstract}

A new mechanism of particle acceleration to ultra high energies,
driven by the rotational slow down of a pulsar (Crab pulsar, for
example), is explored. The rotation, through the time dependent
centrifugal force, can very efficiently excite unstable Langmuir
waves in the e-p plasma of the star magnetosphere via a parametric
process. These waves, then, Landau damp on electrons accelerating
them in the process. The net transfer of energy is optimal when the
wave growth and the Landau damping times are comparable and are both
very short compared to the star rotation time. We show, by detailed
calculations, that these are precisely the conditions for the
parameters of the Crab pulsar. This highly efficient route for
energy transfer allows the electrons in the primary beam to be
catapulted to multiple TeV ($\sim 100$ TeV) and even PeV energy
domain. It is expected that the proposed mechanism may, partially,
unravel the puzzle of the origin of ultra high energy cosmic ray
electrons.

\end{abstract}


%%%%%%%%%%%%%%%%%%%%%%%%%%%%%%%%%%%%%%
\section{Introduction }
%%%%%%%%%%%%%%%%%%%%%%%%%%%%%%%%%%%%%%

If one could devise a mechanism that could convert even a tiny
fraction of the enormous rotational energy of pulsars (spinning
neutron stars) into particle kinetic energy, an outstanding problem
of high energy astrophysics may get closer to a solution. One could,
then, provide, for instance, a possible explanation for what
accelerates cosmic ray electrons to ultra high energies- a regime recently
explored by the H.E.S.S., Pamela and Fermi collaborations
\citep{hessel,pamel,fermel,fermel1}

 Some of the most significant observations and
developments in the realm of high energy cosmic ray electrons
may be summarized as: 1) the $2008$ announcement of the H.E.S.S.
collaboration showing evidence for a substantial steepening in the
energy spectrum of cosmic ray electrons above $600$ GeV
\citep{hessel}, 2) Suggestions for the possible sources  for the
very high energy electrons (VHE electrons) \cite{pamel}: it is
conjectured, for example, that either pulsars or the annihilation of dark
matter particles might be the  source of ultra relativistic energies,
3) the launching of the Fermi spacecraft  into a
near-earth orbit on $11$ June $2008$ - the new instruments on board
detected cosmic-ray electrons up to $1$TeV and confirmed an excess
of VHE leptons in cosmic-rays \citep{fermel,fermel1}, and 4)
Analysis of the observational data of the Fermi telescope on
millisecond pulsars (the surface magnetic field close to $B_{st}\sim
10^{8.5}$G) by \cite{shota} positing that the data may have evidence
of extremely high energy ($\sim 50$TeV ) cosmic ray electrons.

This paper is an attempt to formulate a theoretical
framework for particle acceleration driven by the rotational slow
down of a pulsar, in particular, the Crab pulsar ($B_{st}\sim
7\times 10^{12}$G). The star rotational energy is channeled to the
particles in a two step process - the excitation of Langmuir waves
(via a two-stream instability) in the electron-positron plasma in
the pulsar magnetosphere; and the  damping of Langmuir waves on the
fast electrons accelerating them to even higher energies up to  and beyond $100$
TeV.

All known pulsars are observed with decreasing spinning rates. The
Crab pulsar, for example, rotates with a period $P\approx 0.0332$s
with a characteristic rate of change $\dot{P}\equiv dP/dt\approx
4.21\times 10^{-13}$ss$^{-1}$. The slowing down releases
rotational energy that must transform into some other form. The
slowdown luminosity of the star is given by
$\dot{W}\approx I\Omega\dot{\Omega}$, where $I=2MR_{st}^2/5$ is the
moment of inertia of the pulsar, $R_{st}\approx 10^6$cm is its
radius, $\Omega\equiv 2\pi/P$ is its angular velocity,
$\dot{\Omega}\equiv d\Omega/dt$, and the pulsar mass
$M\sim 1.5\times M_{\odot}$ where $M_{\odot}\approx 2\times
10^{33}$g measures the solar mass. Out of this enormous
output, $\dot{W}\approx 5.5\times 10^{38}$erg/s, only a part
$\sim 2\times 10^{38}$erg/s  is sufficient to power the Crab
Nebula; the rest - almost $60\%$ of the energy budget - is released
via, yet, unknown channels.

In the standard pulsar model \citep{deutsh,gj}, an electrostatic
field uproots particles from the pulsar's surface. These particles
are then accelerated along the magnetic field lines by the nonzero
longitudinal electric field. The electrons, accelerated to relativistic energies, begin to
emit copious curvature radiation. The emitted photons with energies
beyond the pair production threshold
 $2mc^2$ ($m$ is the electron's mass and $c$ is the speed of
light), in turn, create  electron-positron pairs thought the
channel, $\gamma + {\bf B}\rightarrow e^{+}+e^{-}+\gamma'$ (${\bf
B}$ is the dipolar magnetic field). Newly produced pairs are further
accelerated and emit photons; The  cascade lasts till the nascent
pair plasma, especially in regions farthest from the star, is dense
enough to screen out the initial electrostatic field \citep{sturrok,tademaru}.

This height $h$ of the gap region (where the electric field is
nonzero) is, therefore, strictly limited. The limited accelerating
region, in turn, imposes an upper limit on the maximum attainable
electron energy. The potential difference in the gap has been
estimated to be $\Delta V\approx 1.6\times
10^{12}B_{12}^{-1/7}P^{-1/7}$V \citep{rud}, where $B_{12}\equiv
B/10^{12}G\approx 6.7$ is the dimensionless magnetic induction. Thus
the potential difference, predicted by the standard gap model, may boost
the electron Lorentz factors to $\sim 4\times 10^6$ - not
sufficient to explain the observed very high energy (VHE) radiation.
Figure 1 shows the schematic representation of the distribution
function of plasma particles in standard models of the pulsar
magnetosphere. The primary beam particles are shown by the narrow
shaped area with the Lorentz factor $\gamma_b$. The wider region
represents the distribution of secondary electron-positron pairs,
characterized by a corresponding Lorentz factor, $\gamma_p$ when the
distribution function has a maximum value.

Several mechanisms have been proposed to, somehow, enlarge the gap
and boost up energy accumulation in this zone. In
particular, \cite{usov} proposed the so-called intermediate
positronium formation mechanism, and \cite{arons} studied the field lines
supposedly curved towards the rotation and rectifying under certain
conditions. Invoking  general relativity,
\cite{muslimov} considered the role of Kerr metric in the
creation of the additional electric field. In all these attempts,
the gap size does increase but not enough to account for the observations.

Equally noteworthy are the attempts to understand high energy
acceleration in terms of the Fermi acceleration mechanism
\citep{blfermi,blostr}. The Fermi-type acceleration is found to be
efficient when the seed population of pre-accelerated electrons
possesses quite high Lorentz factors \citep{rm00}. The problem of
reaching such high Lorentz factors, still, remains unresolved.

It is in this backdrop - searching for a mechanism for particle
acceleration- that we turn to rapidly rotating pulsars. The
challenge will lie in uncovering a "new" mechanism  for  pumping out
and harnessing energy from the central engine.

We explore, here, a highly efficient mechanism for
transferring energy from pulsar rotation to particle kinetic energy.
The invoked mechanism, first suggested (in a limited context) in
\citep{incr}, consists of two principal steps: 1) The time dependent
centrifugal force, imparted to the relativistic plasma particles
from star rotation, parametrically, drives Langmuir waves,  and 2)
These waves damp on very energetic electrons to accelerate them even
to higher energies. In order to appreciate the workings of the
mechanism, the following short review of the standard picture of the
composition of the pulsar magnetosphere may be necessary.

When acting on relativistic particles, the centrifugal force induces a periodic motion in time
\citep{incr}, and, as a result,  ends up fuelling a parametric instability that may
efficiently pump energy from rotation into Langmuir waves. The
possibility of parametric resonance in an electron plasma was first
discussed by \cite{mt78,m78}. It is worth noting that centrifugal
effects strongly influence particle acceleration and MHD processes
in the magnetospheres of pulsars \citep{forcefree,or09} and AGN
\citep{osm7}.

We will begin, here, by a radical and far-reaching generalization of the mechanism,
first introduced in \cite{incr}. The refashioned model, applied to the Crab pulsar,
departs from the very simplified approach of \cite{incr}
that examined  only a narrow subclass of centrifugally
accelerated relativistic electrons- the ones with
radial velocity $\upsilon_0(t)\approx c\cos\Omega t$ \citep{mr94}.
We, instead, extend the initial particle velocity profile by letting
 $\upsilon_0(t)\approx c\cos\left(\Omega t
+\phi\right)$ with $\phi\neq 0$, the general solution allowed by the centrifugal force.
 It is somewhat unexpected and  rather amazing that insisting on a nonzero initial phase
spawns a rather strong instability.

After this long introduction, we
work out, in Section 2, the theory of centrifugally induced Langmuir waves. In
section 3, we apply the model to the Crab pulsar and in section 4,
we summarize our results.

\begin{figure}
  \centering {\includegraphics[width=7cm]{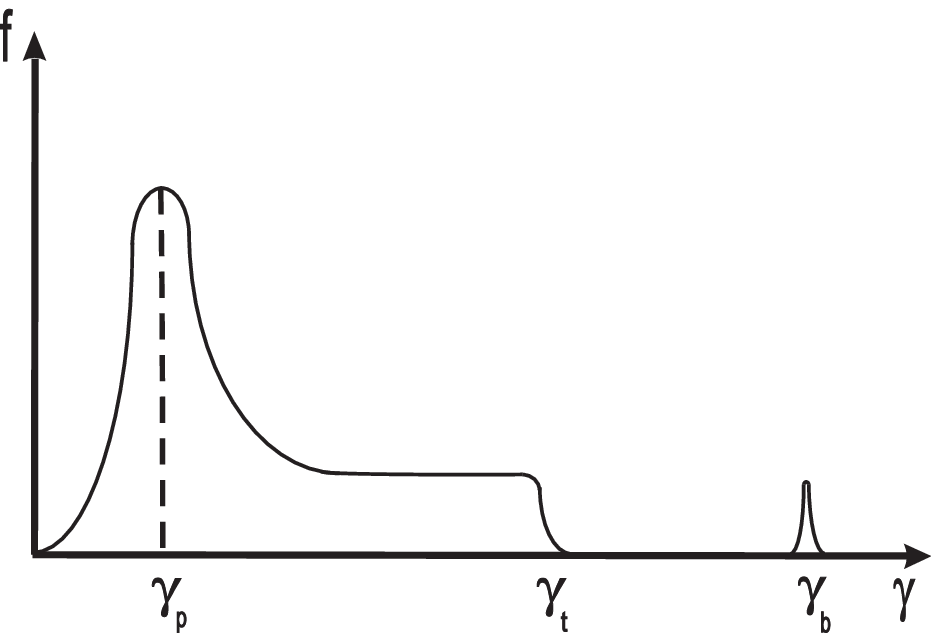}}
  \caption{The distribution function versus the Lorentz factor.
  As is clear from the plot, the function consists of two parts: the first -
  wider region concerns the plasma component corresponding
  to the cascade processes of pair creation and the second one characterizes the primary
  Goldreich-Julian beam electrons.}\label{fig0}
\end{figure}

%%%%%%%%%%%%%%%%%%%%%%%%%%%%%%%%%%%%%%%%%%%
\section[]{Centrifugally Driven Langmuir Waves in the Pulsar magnetosphere}
%%%%%%%%%%%%%%%%%%%%%%%%%%%%%%%%%%%%%%%%%%%

To develop a model for the centrifugally
induced parametric Langmuir instability, we
will assume that the magnetic field lines are almost
rectilinear- a good approximation for distances less than
the curvature radius of the field lines. Then in the rotating frame
of straight field lines co-rotating with the pulsar, the metric is
given by
\begin{equation}
\label{metric} ds^2 =
c^2\left(1-\frac{\Omega^2r^2}{c^2}\right)dt^2-dr^2.
\end{equation}
In the $1+1$ formalism \citep{membran} for a single particle, the
particle equation of motion, in the rotating frame of reference,
reads
\begin{equation}
\label{eul1} \frac{d{\bf p}_{i}}{d\tau}= \gamma_{i}{\bf
g}+\frac{e_{i}}{m}\left({\bf E}+ \frac{1}{c}\bf v_{i}\times\bf
B\right),
\end{equation}
where ${i}=e,p$ is the species index, $e_{i}$ is the charge of the
corresponding particle, $d\tau\equiv\xi dt$,
$\xi\equiv\left(1-\Omega^2r^2/c^2\right)^{1/2}$ is the so-called
time lapse function, ${\bf g}\equiv -{\bf\nabla}\xi/\xi$ is the
gravitational potential, $\gamma_{i}\equiv\left(1-{\bf
V}_{i}^2/c^2\right)^{-1/2}$ is the Lorentz factor, and ${\bf p}_{i}$
and ${\bf V}_{i}$ are, respectively, the dimensionless momentum
(${\bf p}_{i}\rightarrow {\bf p}_{i}/m$) and velocity. The first
term on the righthand side of the equation represents the
centrifugal force, and plays a crucial role in our model.

To transport the equation of motion into the laboratory frame, we use the well
known relation $\gamma = \xi\gamma'$ connecting Lorentz factors in the
rotating ($\gamma$) and laboratory ($\gamma'$) frames. The
identity $d/d\tau\equiv\partial/\partial (\xi\partial t) +
\left({\bf v\cdot\nabla}\right)$ takes Eq. (\ref{eul1}) into the laboratory frame
\begin{eqnarray}
\label{eul2} \frac{\partial{\bf p}_{i}}{\partial t}+({\bf
v_{i}\cdot\nabla)p}_{i}= \nonumber \\
=-c^2\gamma_{i}\xi{\bf\nabla}\xi+\frac{e_{i}}{m}\left({\bf E}+
\frac{1}{c}\bf v_{i}\times\bf B\right),
\end{eqnarray}
where we have omitted the primes for notational simplicity. Note
that the first term, originating in the relativistic effects of
co-rotation, has a singular behavior on the light cylinder surface
(a hypothetical zone, where the linear velocity of rotation exactly
equals the speed of light). Throughout the area of interest, the
overall effect of the term $c^2\gamma_{i}\xi{\bf\nabla}\xi$  will be
significant.

Equation (\ref{eul2}) is clearly fluid-like. In order to motivate
this change in appearance, we need to make a digression. The plasma
we plan to describe is  really a kinetic fluid, the distribution
function covers a wide range of $\gamma$ (energy) and as will later
insist, a variable phase. A proper detailed description, then, is rather
involved and will require numerical modeling. In this paper,
however, we plan to extract our main results in a highly reduced
description. We will assume that the plasma is multi-stream, each
stream with a characteristic $\gamma$ and a characteristic phase. Then we will
solve the linear "interacting" dynamics of two such streams and
demonstrate how a very interesting manifestation of the two stream
instability takes place; the phase difference between the streams
will emerge as the driver. Each stream, of course, is an
electron-positron plasma. Equation (\ref{eul2}), then, is to be
viewed as equation of motion of the specie i(electron or positron)
in  a given stream, and along with the continuity equation
\begin{equation}
\label{cont} \frac{\partial n_{i}}{\partial t}+{\bf
\nabla}\cdot\left(n_{i}{\bf v_{i}}\right)=0
\end{equation}
and the Poisson equation
\begin{equation}
\label{pois} {\bf \nabla\cdot E}=4\pi\sum_{i}e_{i}n_{i}.
\end{equation}
completes the dynamics.

The centrifugal force \citep{mr94} is the cause of  equilibrium or
background velocity for the stream particles. The electromagnetic
terms do not contribute (to the equilibrium) because of the
frozen-in condition ${\bf E_0 + v_{0i}\times B_0 = 0}$. In
equilibrium, Eq. (\ref{eul2}) reads \citep{mr94}
\begin{equation}
\label{eul0}
\frac{d^2r}{dt^2}=\frac{\Omega^2r}{1-\frac{\Omega^2r^2}{c^2}}\left[1-\frac{\Omega^2r^2}{c^2}
-\frac{2}{c^2}\left(\frac{dr}{dt}\right)^2\right].
\end{equation}
For highly relativistic particles ($\gamma\gg 1$), Eq. (\ref{eul0}) yields the simple periodic solution,
\begin{equation}
\label{rt} r(t) \approx \frac{V_0}{\Omega}\sin\left(\Omega t +
\phi\right),
\end{equation}
\begin{equation}
\label{vt} \upsilon_0(t) \approx V_0\cos\left(\Omega t +
\phi\right).
\end{equation}
The phase $\phi$, necessary to insure a general solution of  Eq.
(\ref{eul0}), was taken to be zero in previous work \citep{incr}.
This was a very  serious shortcoming because the  zero phase
solution applies only to a very small subset of particles. The nonzero
phase will insure the inclusion of all particles in the dynamics.
Consequently, the  instability could feed on much of the population of the
magnetospheric plasma resulting in very large growth rates
 .

The orbits (\ref{rt},\ref{vt}), along with the frozen-in condition, provide the equilibrium input for the
linear stability analysis. Since the calculated equilibrium is time dependent, the standard time -Fourier anaylsis
will not pertain- the solution of the stability problem will be more involved and nuanced.

One begins with the standard splitting of  the physical variables
 $\Psi = (n,{\bf v,p,E,})$ into equilibrium and perturbed parts
\begin{equation}
\label{psi} \Psi \approx\Psi^0+\Psi^1,
\end{equation}
The next step is somewhat different; the linear perturbations are
expanded as
 \begin{equation}
\label{psi1} \Psi^1(t,{\bf r}) \propto\Psi^1(t)\exp\left[i({\bf
kr})\right],
\end{equation}
where both the amplitude and the exponent are functions of time. To
keep the notation simple, we will drop the superscript $1$  labelling
perturbed quantities.

In the linearized version of Eqs. (\ref{eul2}-\ref{pois}):
\begin{equation}
\label{eul3} \frac{\partial p_{i}}{\partial
t}+ik\upsilon_0p_{i}=
\upsilon_0\Omega^2rp_{i}+\frac{e_{i}}{m}E,
\end{equation}
\begin{equation}
\label{cont1} \frac{\partial n_{i}}{\partial
t}+ik\upsilon_{{i}0}n_{i}, + ikn_{{i}0}\upsilon_{i}=0,
\end{equation}
\begin{equation}
\label{pois1} ikE=4\pi\sum_{i}e_{\beta}n_{i},
\end{equation}
$E$ is the electrostatic field, and i is the species index. Notice
that we have retained only the radial component of the linearized
momentum equation, $ p_{i}=\widehat {{\bf e}_r}\cdot {\bf p}_i$ and
$k=\widehat {{\bf e}_r}\cdot {\bf k}$.

The electron-positron dynamics can be combined in terms of two fluid variables:, the relative velocity
${\bf v\equiv v_e-v_p}$, the differential density$n\equiv n_e-n_p$,  the average fluid  velocity ${\bf v_e^0
\approx v_p^0\equiv v^0}$, and  the mean density, $n_e^0 = n_p^0\equiv n^0$. Thus for the two stream model
(labelled by $\beta$=1,2), the linear system (\ref{eul3}-\ref{pois1}) takes the form,

\begin{equation}
\label{eul4} \frac{\partial p_{\beta}}{\partial
t}+ik\upsilon_{\beta0}p_{\beta}=
\upsilon_{\beta0}\Omega^2rp_{\beta}+\frac{2e}{m}E,
\end{equation}
\begin{equation}
\label{cont2} \frac{\partial n_{\beta}}{\partial
t}+ik\upsilon_{{\beta}0}n_{\beta}, +
ikn_{{\beta}0}\upsilon_{\beta}=0
\end{equation}
\begin{equation}
\label{pois2} ikE=4\pi e\sum_{\beta}n_{\beta0};
\end{equation}
the two streams being coupled through the Poisson equation.
We remind the reader that  the two-stream model was invoked, very crudely,
to capture the essence of the vast spectrum in energy and initial orbit phase available to the plasma particles.
Each stream carries its own $\gamma$ and phase $\phi$.

Equations (\ref{eul4}-\ref{pois2}), due to the time dependence of the orbits ($r$, and $v_0$), represent  a set of non-autonomous differential equations. The goal is to determine if an arbitrary perturbation will grow in time. We will follow two complementary methodologies to establish the existence of an instability and to determine the growth characteristics;
an exact numerical solution and an approximate analytical estimate via a quasi-modal approach.

We convert (\ref{eul4}-\ref{pois2}) into two coupled equations
\begin{equation}
\label{ME1}
\frac{d^2N_1}{dt^2}+{\omega_1}^2 N_1= -{\omega_1}^2 N_2 e^{i \chi}
\end{equation}
\begin{equation}
\label{ME2}
\frac{d^2N_2}{dt^2}+{\omega_2}^2 N_2= -{\omega_2}^2 N_1 e^{-i \chi}
\end{equation}
in terms of the perturbed "densities'
\begin{equation}
\label{var}
N_{\beta}= n_{\beta}e^{\frac{ic
k}{\Omega}\sin\left(\Omega t + \phi_{\beta}\right)},
\end{equation}
where we have replaced $V_0$ by $c$. The frequencies $\omega_{1,2}\equiv\sqrt{8\pi e^2n_{1,2}/m\gamma_{1,2}^3}$ are, respectively the effective relativistic plasma frequencies (for longitudinal modes) corresponding to the two streams, and
\begin{equation}
\label{exp1}
\chi=\frac{ck}{\Omega}[\sin (\Omega t + \phi_1)-\sin (\Omega t + \phi_2)]= b \cos (\Omega t+\phi_+ )
\end{equation}
with $2 \phi_{\pm}= \phi_1\pm\phi_2$, and
\begin{equation}
\label{exp2}
b=2 \frac{ck}{\Omega} \sin\phi_-
\end{equation}
In deriving Eqs.(\ref{ME1}-\ref{ME2}), we have invoked
$(ck/\Omega)>>1$ to drop the first term on the right hand side of
Eq. (\ref{eul4}). This  parameter will turn out to be approximately
the ratio of the Langmuir frequency  to the star rotation frequency
and will lie in the range $10^6-10^{10}$.

The evolution Eqs.(\ref{ME1}-\ref{ME2}) with definitions
(\ref{exp1}-\ref{exp2})  have the following features:

1) The evolution of the two streams is coupled.

2) The time dependence of the coefficients is fully contained in
$\chi$.

3) $\chi$ depends on both the sum and difference of the phases
associated with the two streams.

4) The dependence on the phase sum $2\phi_+=\phi_1+\phi_2$ is
trivial. In fact it can be absorbed in a redefinition of time
variable; $ \phi_+$ will be dropped from now on.

5) The phase difference $ \phi_-$, however, is a fundamental
parameter of the two-strem system and determines the magnitude of
$\chi$. For $ \phi_-=0$, $\chi\approx 0$,
Eqs.(\ref{ME1}-\ref{ME2}) reduce to an autonomous system
with constant coefficients and lead to the modal
dispersion relation ${\omega}^2= {\omega_1}^2+{\omega_2}^2$. The two
streams  collapse to one since $ \phi_-$, the "measure" of their
difference, has gone to zero. Strictly speaking, this is only an
approximation, since in Eq. (\ref{exp1}) we have assumed $V_{0\beta}
= c$, whereas $V_{0\beta} \approx
c\left(1-\frac{1}{2\gamma_{0\beta}}\right)$. Without this approximation, we will get a nonzero
growth rate even for $\phi_-=0$). However, this vestigial growth rate is
many orders of magnitude smaller than what we  will obtain for finite $ \phi_-$.

We would reassert that introducing the initial phase is a
fundamental departure from earlier work. It is the phase difference that
differentiates the two streams and sets the stage for a strong two stream instability.

We will first perform a time Fourier analysis to search for a
growing quasi mode (for non autonomous systems there are no fourier
modes). We must manipulate Eqs.(\ref{ME1}, \ref{ME2}) to yield
\begin{equation}
\label{ME3} \frac{d^2N_2}{dt^2}=\frac{{\omega_2}^2}{{\omega_1}^2}
e^{-i \chi} \frac{d^2N_2}{dt^2}.
\end{equation}
We, then, use the Bessel identity
\begin{equation}
\label{iden} e^{\pm ib\sin y} = \sum_\mu J_{\mu}(b)e^{\pm i\mu y},
\end{equation}
and take the Fourier transform of Eqs.(\ref{ME1}-\ref{ME3}) to derive the relation
\begin{equation}
\label{disp1}
( \omega^2 -\omega_1^2)N_1(\omega)=\omega_2^2\sum_{\mu\nu}Q_{\mu}^{+}Q_{\nu}^{-}\frac{(\omega+(-\mu+\nu)\Omega)^2}{(\omega-\mu\Omega)^2}N_1(\omega+(-\mu+\nu)\Omega )
\end{equation}
where
 $$ Q_{\mu}^\pm=e^{\pm i\mu \frac{\pi}{2}} J_{\mu}(b)$$

Equation (\ref{disp1}) is clearly not a dispersion relation, each
Fourier component (labeled by $\omega$) is connected to all other
components. In spite of the formidable double sum on the right hand
side, we can still extract useful information that will establish
the possibility of the growth of perturbations. An indicative "
dispersion relation "relation may be obtained by keeping terms with
$\nu=\mu$
 \begin{equation}
 \label{disp2}
 \omega^2 -\omega_1^2 - \omega_2^2  J_0^2(b)= \omega_2^2 \sum_{\mu} J_{\mu}^{2}(b)
 \frac{\omega^2}{(\omega-\mu\Omega)^2}.
 \end{equation}

We make  further analytical  progress by simply keeping a single resonant term defined by
 the harmonic of the star rotation frequency that is in resonance
with the real part of the mode frequency, $\omega_r=\mu_{res}\Omega$
where $\omega=\omega_r +\Delta$ ( $\Delta<<\omega_r$). After some
straightforward algebra, one can derive

 \begin{equation}
 \label{disp3}
 \Delta^3=\frac{\omega_r {\omega_2}^2 {J_{\mu_{res}}(b)}^2}{2},
 \end{equation}
implying an imaginary part- growth rate:
\begin{equation}
 \label{grow1}
 \Gamma= \frac{\sqrt3}{2}\left (\frac{\omega_r {\omega_2}^2}{2}\right)^{\frac{1}{3}}
 {J_{\mu_{res}}(b)}^{\frac{2}{3}}.
\end{equation}
The preceding analytical calculation, though
simple and revealing, should be taken only as an indicator that the
Langmuir waves (charge separation perturbations) can grow in time
transferring energy from pulsar rotation to the electric field. As
remarked earlier the instability is an appropriate version of the
two-stream instability in which both streams move with speeds close
to $c$, but whose constituent particles start with different initial
orbits-the difference manifesting itself through the phase factor
$\phi_-$.

The expression (\ref{grow1}) for the growth rate tells us that,
apart from its dependence on $\omega_1 $ and  $\omega_2$, the growth
rate is controlled by the Bessel  function
$J_{\mu_{res}}(b)=J_{\mu_{res}}((2 ck/ \Omega) \sin\phi_-)$ of very
high order $\mu_{res}=\omega/\Omega\gg 10^4$. Bessel functions of
such high order are vanishingly small  unless the argument $(2 ck/
\Omega) \sin\phi_-)\approx\mu_{res}$. For $\phi_-$ approaching zero, the growth
rate becomes vanishingly small \citep{incr}.

We could readily make analytical estimates for the growth rate of
the rotationally driven Langmuir waves for a typical pulsar. However
we must bear in mind that these estimates can be merely suggestive
since the fundamental dynamics is non-autonomous in time and a given
Fourier mode cannot capture the time behavior. It is particularly so
because $\mu_{res}$ is so large that  a whole band of harmonics will
contribute to the sum (\ref{disp1}) even if mode coupling were
neglected. However, the evolution of each mode is governed by others
and the time asymptotic response does depend on the coupling of
Fourier modes- we must deal with the entire wave packet.

Fortunately the system of Eqs.(\ref{ME1}-\ref{ME2}), with periodic
coefficients, is readily amenable to a simple numerical solution for
the densities $N_1$ , $N_2$ (and the electric field, $E= N_1-N_2
e^{i \chi}$). Differential equations with periodic coefficients
(Mathieu equation being the simplest example) are known to display
parametric instabilities in well defined regimes. Mathematica
solutions demonstrate that model Eqs.(\ref{ME1}-\ref{ME2}) are,
indeed, parametrically unstable and acquire a hefty growth rate when
$b\approx\omega/\Omega$ (this behavior was predicted by the
analytical dispersion relation). It turns out the time asymptotic
behavior, for cases of interest, does settle into a pattern that
growth rate of the instability can be read off from the time
evolution graph. We shall soon see that rotation (centrifugally)
driven Langmuir waves can grow at a very fast rate.

%%%%%%%%%%%%%%%%%%%%%%%%%%%%%%%%%%%%%%%%%%%%%%%%%%%%%%%%%%%%%%%%%
\section{Discussion}
%%%%%%%%%%%%%%%%%%%%%%%%%%%%%%%%%%%%%%%%%%%%%%%%%%%%%%%%%%%%%%%%%%%%%%%%%%%%

We shall now explore the detailed characteristic of the centrifugally driven
Langmuir waves for the crab pulsar. Our eventual  aim is to put together a mechanism for delivering
the rotational slow down energy to particle acceleration. It will happen in two major
steps: (I) in the first stage, by means of the centrifugally induced
parametric instability, the pulsar's rotation energy is transferred
to the electrostatic waves; (II) and in the second stage the amplified
electrostatic modes undergoes  Landau damping on high energy electrons causing efficient particle acceleration.

%%%%%%%%%%%%%%%%%%%%%%%%%%%%%%%%%%%%%%%%%%%%%%%%%%%%%%%%%%%%%%%%%
\subsection{Centrifugally induced Langmuir waves}
%%%%%%%%%%%%%%%%%%%%%%%%%%%%%%%%%%%%%%%%%%%%%%%%%%%%%%%%%%%%%%%%%%%%%%%%%%%%

In order to make reasonable estimates for the instability growth
rate, we recapture some details of our model. What we are studying
is the instability that arises in a highly relativistic
electron-positron plasma in the pulsar magnetosphere. We have
modeled  the plasma in terms of two distinct streams distinguished
by their $\gamma$ factors  and the initial orbit phase (we have
assumed the speed of both streams to be c). Let the two streams have
quite different $\gamma$s with $\gamma_1<\gamma_2$. The frequencies
$\omega_{1,2}$ are calculated from the stream densities and
$\gamma_{1,2}$. Assuming that energy is uniformly distributed
throughout the magnetospheric plasma, we may approximate
$n_1\gamma_1\approx n_2\gamma_2\approx n_b\gamma_b$, where $n_b =
n_{_{GJ}} = \Omega B/(2\pi ec\xi^{2})$ is the Goldreich-Julian
number density \citep{rud}. The magnetic field (on the light
cylinder zone) $B = B_{st}(R_{st}/r_0)^3$ is related to
$B_{st}\approx 6.7\times 10^{12}$, the magnetic field close to the
neutron star's surface. With these density relations, we conclude
that one of the parameters  for the numerical solution,
$\alpha={\omega_2}^2/{\omega_1}^2={\gamma_1}^4/\gamma_2^4$ will be
much smaller than unity. The most important determining parameter of
the system, however, is $ w = \omega_1/ \Omega$, the ratio of the
rotation frequency to the mode frequency ($\approx \omega_1$);  the parameter
$w$ will lie, typically, between $10^6-10^{10}$, the former near the light
cylinder and the latter near the pulsar surface. Specifying
$\alpha$, $w$, $b$, and  the initial conditions is sufficient to
seek a numerical solution. The parameter $b$ is set by the wave
number k and the phase difference $\phi_-$. From the general
character of the differential equations, we expect strong parametric
amplification when $b\ge w$.

\begin{figure}
  \centering {\includegraphics[width=7cm]{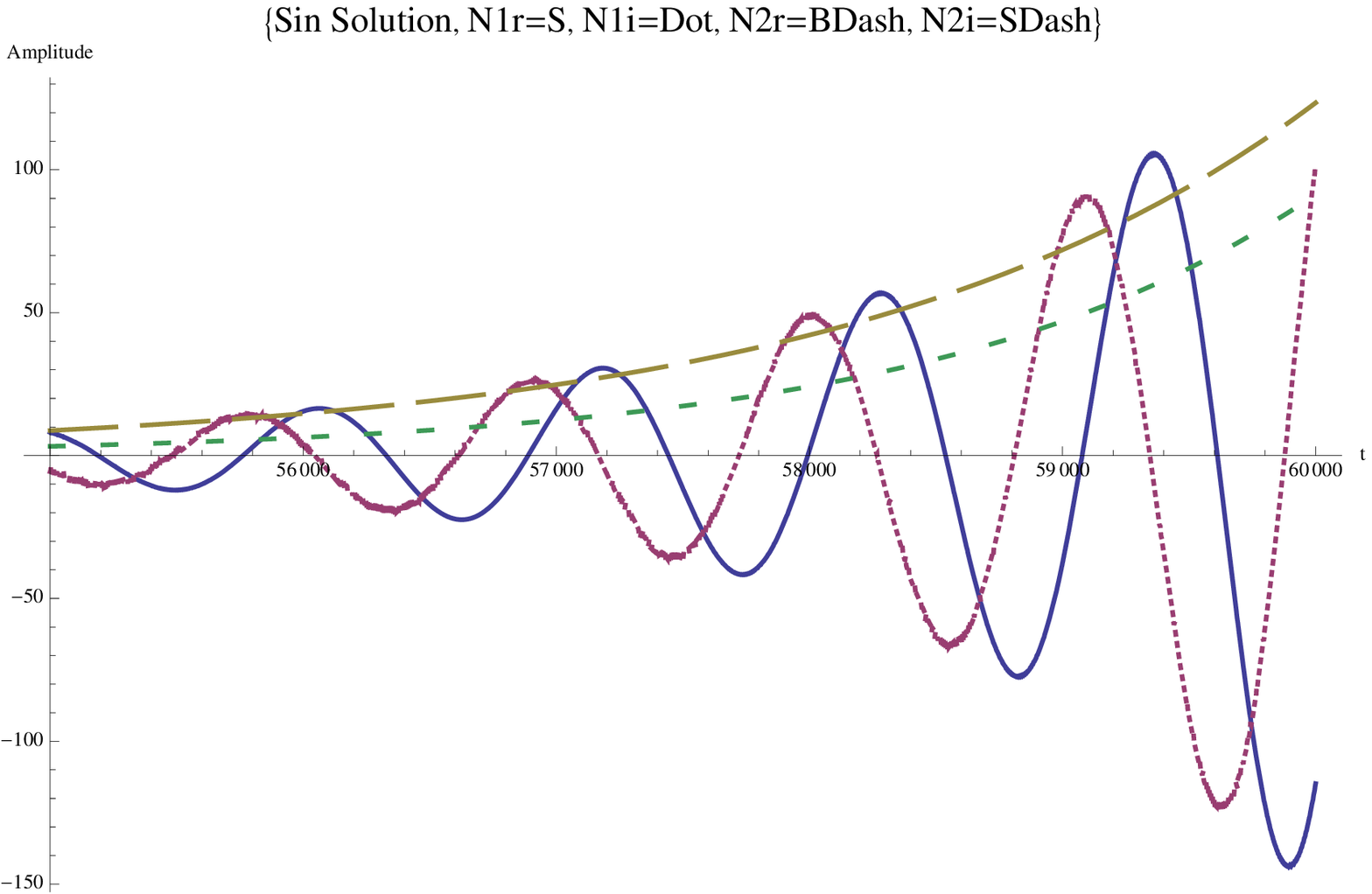}}
  \caption{The temporal behaviour of amplitudes of $ReN_1$ (S), $ImN_1$ (Dot), $ReN_2$ (BDash), $ImN_2$ (SDash).}\label{fig1}
\end{figure}

\begin{figure}
  \centering {\includegraphics[width=7cm]{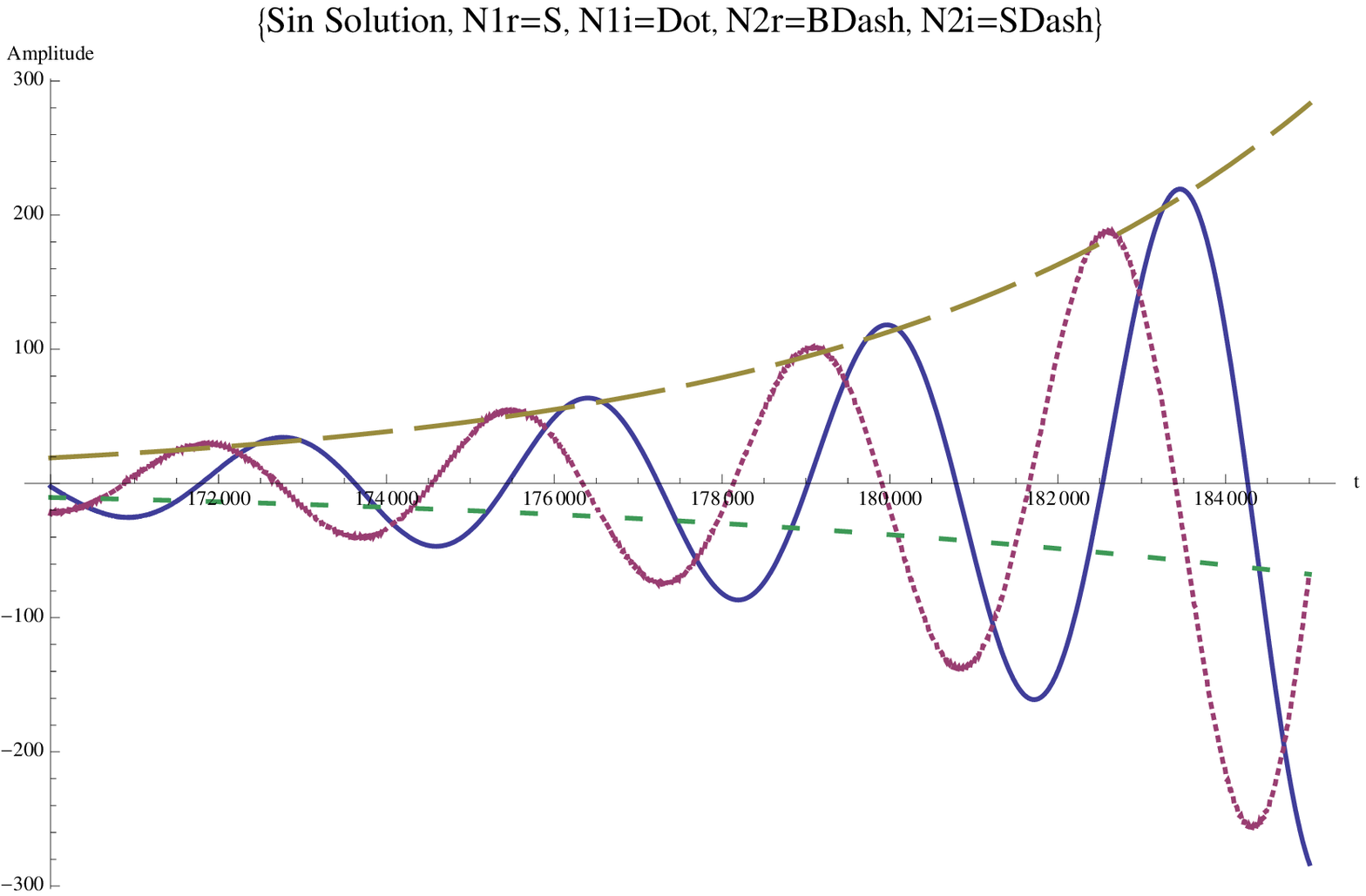}}
  \caption{The temporal behaviour of amplitudes of $ReN_1$ (S), $ImN_1$ (Dot), $ReN_2$ (BDash), $ImN_2$ (SDash).}\label{fig2}
\end{figure}

From our extensive numerical study, we pick up two representative
Mathematica plots (Fig. \ref{fig1} and Fig. \ref{fig2}). In these, the solutions for $ReN_1$, $ImN_I$, $ReN_2$, and $ImN_2$ are plotted as functions of time. Time has been
normalized to the inverse of the frequency $\omega_1$, the
relativistic plasma frequency corresponding to the stream with lower
$\gamma$ and much higher density. The respective parameters for the two figures are: For
Fig. \ref{fig1}, $w=b=10^7$, $\alpha=.01$, and for Fig. \ref{fig2},
$w=b=10^8$, $\alpha=.01$. Since the $x$ axis denotes time measured
in  their respective plasma time (inverse of $\omega_1$)- the
"absolute" time scales in the two figures are different by a factor
$\sqrt{10}$. For each case, we choose initial conditions(t=0) for
what we call the Sin solution: $N_1(0)=0, {N_1}'(0)=1,N_2(0)=0,
{N_2}'(0)= \alpha(\cos{b}-i\sin{b})$; it is easy to verify that
these constitute a consistent solution to the coupled differential
equations in the limit $t\rightarrow 0$.

The principle results of the numerical investigations may be  summarized as:

a) For initial times $t/w <<1$(not displayed in the figures), the
streams are essentially uncoupled and each oscillates at its own
plasma frequency; the frequencies are real. Since $w$ is large, this
stage can be reasonably long. For the two cases displayed, the
period lasts $\approx3.10^4$ plasma periods for case 1, and
$\approx2.10^5$ plasma periods for case 2. In each case the period
is $\approx10^{-3}$ of the star rotation period. Compared to the
star rotation time, the coupling and instability build-up time is
very small for the displayed examples.

b)As long as $b\ll w$, even for large $t/w$, the coupling between
the streams remains small. Consequently, the perturbations
amplitudes remain limited.

c) As $b$ approaches $w$, the two streams begin to interact after
some initial lapse of time and all 4 amplitudes begin to grow. Since
the two streams are assumed to have different intrinsic plasma
frequencies
($\alpha={\omega_2}^2/{\omega_1}^2={\gamma_1}^4/\gamma_2^4<<1$) the
time dependence of ($ReN_1, ImN_1)$, and  ($ReN_2, ImN_2)$ are quite
different. It can be seen from the figures that although ($ReN_1,
ImN_1)$ oscillate (at a new slow frequency, the plasma oscillations
are very fast and ever present-not resolved in the pictures) and
grow, the amplitudes ($ReN_2, ImN_2)$ simply grow. Interestingly
enough, the slow time evolution is exponential as predicted by the
Floquet theory of  differential equations with periodic
coefficients. Thus the amplitude of entire perturbation (not just a
given Fourier mode) exponentiates, and one can define a "global"
growth rate.

d) From the numerical solutions one  can readily extract the the
growth (exponentiation) time. For the solution in Fig. \ref{fig1},
it is $\approx 2000$ plasma times $ \approx 2. 10^{-4}P$, while the
growth rate for the solution in Fig. \ref{fig2} is $\approx 6000$
plasma times $\approx 6. 10^{-5}P$, where $P$ is the pulsar period.
Thus once the instability sets in, its growth rates are very high
compared to the frequency of star rotation.

e) For a given $b$  and $w$, the growth rates go down as
${\alpha}^{1/3}=(\gamma_1/\gamma_2)^{4/3}$, precisely the scaling
implied in the simple analytical estimate (\ref{grow1}). Given that
the  growth rates are quite high, one may conclude that even the
streams  with quite disparate $\gamma$ would produce a hefty
instability.

f) For a moderate and not too small $\alpha$, the primary
determinant of the instability gestation time, and growth rate is
the relative value of $b=2 c k/ \Omega \sin\phi_{-}$ and
$w=\omega_1/\Omega$. Both $\omega_1$ and $\Omega$ are
determined by the pulsar phenomenology and are independent
of the wave characteristics. Thus to bring $b$ to the
range of strong instability, one has to choose
$k\sin\phi_{-}$ to be in the desirable range. In principle, then,
for any finite $\sin\phi_{-}$, one can find a range of $k$ for
strong growth. The centrifugal drive for  Langmuir waves is,
therefore, very robust and effective over a wide spectrum.

g)The two stream model calculation, though far from perfect, is
perfectly capable of capturing the essence of the pumping of
Langmuir waves in the magnetospheric electron-positron plasma by
the star rotation. A more detailed and comprehensive semi-kinetic
model, that treats $\gamma$  and $\phi_-$ as phase space variables,
is certainly recommended. However the current calculation creates a
strong basis for the rapid build-up of rotationally pumped Langmuir
waves (and their associated  longitudinal electric fields) so that
we can confidentlly begin to explore the consequences of these
electric fields for particle acceleration.

%\begin{figure}
%  \resizebox{\hsize}{!}{\includegraphics[angle=1]{Fig.a1}}
%  \caption{The}\label{figaa1}
%\end{figure}

%  I have rewritten upto this point

%By substituting these quantities
%into Eq. (\ref{delta3}) and taking into account the equilibrium
%distribution of energy one can reduce the expression of the
%instability growth rate \citep{abram}
%\begin{equation}
%\label{delta4} \Gamma\approx 1.93\times 10^5\times
%\frac{\gamma_l^{2/3}}{\gamma_h^{4/3}}\times
%\cos^{4/3}\left(0.69\mu_0-\frac{\pi}{4}\right)s^{-1}.
%\end{equation}

%we take into account the fact that the mentioned mechanism is
%extremely efficient on the light cylinder lengthscales. For
%simplicity we consider the development of the instability for
%$r_0=0.7R_{lc}$ (see Eq. (\ref{rt})), where $R_{lc}\equiv c/\Omega$
%is the light cylinder radius.

%In Fig. \ref{fig1} we show the behaviour of the maximum value of the
%dimensionless electrostatic field versus $\gamma_h$. The field is
%normalized by the initial value. The set of parameters is $\gamma_l
%= 20$, $\gamma_b = 10^7$, $P\approx 0.0332$s, $B_{st}\approx
%6.7\times 10^{12}$G. As it is evident from the plot, by means of the
%instability the electrostatic field becomes very big, which means
%that the process of energy pumping from rotation is extremely
%efficient.

%\begin{figure}
%  \resizebox{\hsize}{!}{\includegraphics[angle=0]{fig3.eps}}
%  \caption{The dependence of energy of the cosmic ray electrons versus $\gamma_2$.
%  The set of parameters is $\gamma_1
%= 800$, $\gamma_b = 10^7$, $P\approx 0.0332$s, $B_{st}\approx
%6.7\times 10^{12}$G.}\label{fig3}
%\end{figure}

%%%%%%%%%%%%%%%%%%%%%%%%%%%%%%%%%%%%%%%%%%%%%%%%%%%%%%%%%%%%%%%%%
\subsection{Particle acceleration}
%%%%%%%%%%%%%%%%%%%%%%%%%%%%%%%%%%%%%%%%%%%%%%%%%%%%%%%%%%%%%%%%%

After having established that the rotational energy of the pulsar can be rapidly converted into
charge separation electrostatic fields, we will now  address the next issue of driving fast particles with
these electric fields. Looking for a mechanism to produce super- energetic particles,
we will concentrate on further energizing the most energetic particles in the pulsar atmosphere-
the primary beam electrons that, to begin with, have rather high Lorentz factors as high as $\gamma_b\sim 10^7$ \citep{or09}.

The principal  energy transfer (from waves to particles) channel explored in this
work is the damping of Langmuir waves on electrons (Landau
damping). Landau damping is, primarily, a kinetic process and will need to be appropriately
integrated into what has been, till now, an essentially fluid approach. Even at the cost of repetition,
it helps to emphasize that, in our model,

1) The Langmuir waves are produced by the bulk e-p plasma characterized by
the relatively lower $\gamma$ region of the distribution (Fig.1),

2) while we want these waves to damp on the electron beam that appears as a small blip
in the very high $\gamma$ end of the spectrum (Fig.1)

The "producer" and the expected "consumer" particles, though sharing the same physical space,
are very different in their numbers, and in per particle intrinsic energy.

The study of electrostatic waves in ultrarelativistic plasmas
started prior to the discovery of pulsars \citep{silin,citovich}, but
gained considerable momentum after the discovery \citep{lms,vkm}.
From this set of papers, one learns that (one dimensional) electrostatic waves in
relativistic e-p plasmas allow very simple dispersion in two limits: when the phase
velocity is very high ($\upsilon_{ph}\equiv\omega_r/k\gg c$), and when the phase
speed is very close to the speed of light, $\upsilon_{ph}\approx c$. It is clearly the
second case that is of interest to the energy transfer mechanism. In this limit,
the real part of the dispersion relation takes the form \citep{lms}
 \begin{equation}
 \label{dispr}
 \omega = kc-\beta (k-k_0),
 \end{equation}
where $\beta = \langle\gamma\rangle/\langle 2\gamma^3\rangle$ is a
very small parameter $\langle\rangle$ denotes momentum averaging),
and $k_0$ is the limiting value of $k$ for which the phase speed is
exactly $c$. From Eq. (\ref{dispr}), we infer that  the phase
velocity will be less than the speed of light if $k>k_0$; this
condition defines the regime in which such a wave could strongly
interact with physical particles.

Going back to the  unstable Langmuir waves, described in the previous section,
we recall that the instability growth rates were found to be highest when the parameter
(argument of the Bessel function) $(2 ck/ \Omega) \sin\phi_-$ is close to but moderately greater
than $\mu_{res}$ (the order of the Bessel function). The "resonance" condition,
 \begin{equation}
 \label{reso}
(2 ck/ \Omega) \sin\phi_-\approx g \mu_{res}= g \omega/\Omega
 \end{equation}
where g is number $\ge 1$, implies that the phase velocity of the unstable waves,
$\omega/k=\upsilon_{ph}\approx (2c/g) \sin\phi_-$, is in the correct regime as long as
$2\sin\phi_- /g$ is close to  but less than unity. For g=1, for example, only the waves generated by streams
whose phase difference, $\phi_{-}\le\pi/6$, will be able to accelerate particles; the highest growth
pertain for waves for which $\phi_{-}$ is asymptotically close to $\pi/6$ and the phase velocity of such
waves is asymptotically close to $c$.

Thus the e-p bulk plasma, preferentially, generates waves  with phase velocity asymptotically approaching $c$.
These waves will, preferentially, react with particles that
move very close to the speed of light.  In the pulsar magnetosphere,
the primary beam particles (the narrow blip on the high $\gamma$ end of
the distribution function shown in Fig. \ref{fig0}), belong to the most relativistic class.
The Langmuir waves, therefore, would, preferentially damp on them. Indeed,
Eq. (\ref{dispr}) tells us that  closer the phase velocity to $c$, the smaller the value of
$\beta = \langle\gamma\rangle/\langle 2\gamma^3\rangle$, and, consequently the larger the value of
required $\langle\gamma\rangle$.

To recapitulate the working of the Landau damping mechanism, let us place the
phase velocity of the wave in that part of the primary beam region in which the distribution
function decreases with $\gamma$ ($\partial f/\partial p<0$ or $\partial f/\partial \gamma<0 $).
In this zone the number of
electrons that satisfy $\upsilon_{b}<\upsilon_{ph}$ is more than the
number of electrons with $\upsilon_{b}>\upsilon_{ph}$,  $\upsilon_{b}$ denoting the
electron speed. Less energetic particles will gain energy from the waves, whereas more energetic
electrons will loose energy, but since the number of particles with
$\upsilon_{b}<\upsilon_{ph}$ is more than  that of particles with
$\upsilon_{b}>\upsilon_{ph}$, the  waves will damp with a net energy transfer to the beam electrons.

As an aside, the parametric instability, discussed here, may be different from the one conjured in a
typical laboratory setting. In the pulsar magnetospheric situation, the plasma
frequency exceeds the rotation frequency by many orders of magnitude. The
resonance is possible only for higher order harmonics and  with
phase velocity slightly less than the speed of light.

From the relativistic plasma literature \citep{vkm},  we can read off the estimated  characteristic damping rate
\citep{vkm}
\begin{equation}
\label{dampr}  \Gamma_{LD}= \frac{n\gamma\omega_p}{n_1\gamma_1^{5/2}},
\end{equation}
where $n$, $\gamma$ and $\omega_p=\sqrt{4\pi e^2 n/m}$ are
respectively the number density, the Lorentz factor and the plasma
frequency of the specie on which the damping occurs.

If the instability growth rates, and  the Landau damping rates were very disparate,
there will be little effective energy transfer from waves to the particles. On the other hand, if the
damping rates were far in excess of the growth rates, the waves will
not grow much, again resulting in very little transfer from the star
rotation to the waves. The most optimum scenario for an overall
efficient energy pumping/transfer system is  realized when the
instability growth and Landau damping rates are large and
comparable, $\Gamma\sim\Gamma_{LD}$.

Since it is the electron-positron plasma that supports and feeds the two-stream instability,
the corresponding characteristic energy  in the unstable waves is of
the order of the energy accumulated (transferred from star rotation) in the plasma components.
Remembering that the number density associated with the e-p-plasma far exceeds that of the
beam component (see Fig.1), the energy gained per beam electron could be extremely large if
a substantial fraction of  total ``available" energy in plasma components were transferred via the
electrostatic fields.

The condition $\Gamma\sim\Gamma_{LD}$ is satisfied, for instance, by the choice
$\gamma_2=4700$, $\gamma_1\approx 800$ for the two streams. The resulting
instability growth rate is of the order of $10^4$s$^{-1}$; the corresponding growth period
of $10^{-4}s$ is several orders of magnitude less than the pulsar
rotation period. The  rate of energy transfer to the waves is extremely fast-
the processes under consideration are extremely efficient.

Let us  probe  further into the details of energy transfer from the star rotation
into the plasma particles. In the local frame of reference the
particles are forced to slide along the magnetic field lines by
means of the centrifugal force. Viewed in the laboratory frame, the
reaction force $F_{reac}\approx2mc\Omega\xi (r)^{-3}$ \citep{grg}, driving the particles,
becomes infinite on the light cylinder surface. This is a natural result- to preserve
rigid rotation, the  particle velocity, in this region, must exactly
equal the speed of light. Note also that the radial
velocity tends to zero, because on the light cylinder  the
particles can only rotate with the linear velocity $c$. It is clear
that the maximum energy gained from the rotator can be estimated as
the work done by the reaction force. For all particles involved in the process, the energy
gain may be estimated as ($n_1\gg n_2$)
\begin{equation}
\label{work}  W\approx n_1\delta VF_{reac}\delta r,
\end{equation}
where $\delta V$ is the volume in which the pumping takes place and
$\delta r\sim c/\Gamma$ is the corresponding lengthscale.

Since the aforementioned work done is transferred to the beam electrons
in the same volume, the energy of a given beam particle $\epsilon$
can be approximately calculated by equating $W$ with the total energy gained by
the accelerated beam particles, $n_b\delta V\epsilon$,
\begin{equation}
\label{en}  \epsilon\approx \frac{n_1F_{reac}\delta r}{n_b}.
\end{equation}
One can readily show that, in the vicinity of the light cylinder
surface ($\xi\sim 10^{-2}$), and for typical magnetospheric parameters,
$n_1\approx 2.5\times 10^{11}$cm$^{-3}$, $n_b=n_{GJ}\approx 2\times
10^{7}$cm$^{-3}$, $\epsilon$, the energy  acquired by a  beam
electron could get to be as very high- $100s$ of TeVs or higher.

Because the highly relativistic particles will, inevitably, loose energy due to radiation,
one must investigate how
radiation will affect overall energy transfer. Could radiation, for example, put a stringent limit on the maximum energy acquired?

For highly relativistic electrons and photons with
$\epsilon\epsilon_{ph}/(m^2c^4)>1$
($\epsilon_{ph}$ is photon energy), the inverse
Compton mechanism  operates in  the Klein-Nishina (KN) regime\citep{Lightman}.
Energy emitted per particle per second is, then, given by the
approximate expression $P_{KN}\propto \pi
r_e^2m^2c^5n_{ph}(\epsilon_{ph})\mid ln\left(4e\epsilon/m^2c^4- 11/4
\right)\mid$ \citep{KN} where $r_e$ is the classical
electron radius. The corresponding time scale of the process
is $t_{KN}\sim \epsilon /P_{KN}$.  Using the approximate photon number density,
$n_{ph}(\epsilon_{ph})\approx L/4\pi R_{lc}^2c\epsilon_{ph}$
($R_{lc}\equiv c/\Omega$ is the the light cylinder radius), and
taking into account that for the Crab pulsar, $L\sim 10^{36}$erg/s,
$\epsilon_{ph}\sim 100$GeV, it is straightforward to show that if
$\epsilon=100$TeV (for instance), the aforementioned timescale
exceeds the typical instability time scale ($t_{in}\sim 1/\Gamma$)
by many orders of magnitude. Thus Compton cooling is very slow
compared to the wave energy transfer time. The KN time scale goes up
with the particle energy implying that at higher energies, the
inverse Compton mechanism in the KN channel is tot slow to impose any
constraints on the maximum attainable electron energy.

The next possible loss mechanism for relativistic particles, moving in magnetic field,
is the synchrotron emission with the corresponding power $P_{syn}\approx
 2e^2\omega_B^2\gamma^2/3c$ \citep{Lightman}, where
$\omega_B\equiv eB/(mc)$ is the cyclotron frequency. One can
readily show that the electrons leaving the gap with a
$\gamma\sim 10^{6}$ will undergo so efficient synchrotron cooling
that the corresponding time scale, $t_{syn}\sim \gamma mc^2/P_{syn}$
will be of the order of $10^{-21}$s. Thus, shortly after
the particles leave the gap, they radiate their transverse momentum, and
very soon transit to the ground Landau state. After that, zipping only along the
field lines, the electrons will reach the light cylinder zone in due course of time.
It is precisely the region, where the Langmuir waves, always propagating along
the field lines, are excited. Therefore the wave interaction with particles will not
cause pitch angle scattering, efficiently suppressing the synchrotron mechanism.
Quasi linear diffusion, another possible source for imparting a pitch angle,
 \citep{lmm}, also  does not operate because the required condition for diffusion
$\omega_B>\omega_{1,2}$, is violated for extremely energetic plasmas
(for plasmas with energy density exceeding that of magnetic field).
The synchrotron mechanism, therefore, is not expected to impose any
constraints on the maximum attainable energies.

How about the curvature radiation emitted by particles moving along the curved magnetic
field lines? High energy electrons with energies in ($100s$ of TeVs), i.e, with  particle energy density
exceeding  magnetic energy density by many orders of magnitude,
move, practically, in a straight line. This in
turn excludes the curvature radiation from interfering with our energy transfer mechanism. As
in previous cases, curvature radiation "does" not impose significant limits on $\epsilon$.

The absolute upper bound on the energy that an electron may acquire,
may be estimated as follows: The total power budget, the slowdown
luminosity, is $5.5\times 10^{38}$erg/s, about $40\%$ of which is
radiated away by the Crab nebula. If the remaining $60\%$), $\Delta
L\sim 3.5\times 10^{38}$erg/s, were fully used up (via the
mechanism proposed in this paper) to accelerate electrons to ultra
high energies,  a beam electron  could emerge with an energy
\begin{equation}
\label{enmax}  \epsilon\approx \frac{\Delta L}{4\pi\eta
R_{lc}^2n_{GJ}c}\approx 1\frac{PeV}{\eta},
\end{equation}
where $\eta<1$ is the fraction of particles involved in the process.
Since the particles involved in the acceleration-radiation
mechanisms, are from the region of open magnetic field lines, it is
natural to assume that $\eta\ll 1$; this bunch of particles could be
boosted up to multi PeV range. This estimate should be taken for
what it is- an absolute upper bound. The most important message is
that even if a very small percent of the total available energy
could be transferred to fast electrons, the pulsar could be a source
of  ultra relativistic electrons with energies in the range of
$100s$ of Tevs.

It is clear from the expression of $F_{reac}$ that the force
increases asymptotically as we approach the light cylinder zone
$\xi\rightarrow 0$. However, the projected singularity at the light
cylinder surface may be circumvented by a recently conjectured
mechanism: By means of a curvature drift instability, the magnetic
field lines twist' and, lagging behind the rotation, lead to the
force-free dynamics of particles. The rigid rotation and the
subsequent violation of the causality principle \citep{forcefree}
is, thus, prevented. As a result, the process will, then, stop due
to the aforementioned instability.

\section{Summary}
%%%%%%%%%%%%%%%%%%%%%%%%%%%%%%%%%%%%%%%%%%%%%%%%%%%%%%%%%%%%%%%%%%%%%%%%%%%%%%%%%%%%%
\begin{enumerate}

 \item We have constructed a simple but nontrivial theoretical framework  to explore the possibility of particle acceleration driven by the rotational slow down of a pulsar. The first step in the proposed mechanism is the conversion of the rotation
imparted centrifugal particle energy, via a "two stream instability", to electrostatic
Langmuir waves in the electron - positron plasma residing in the pulsar magnetosphere.
Landau damping of these centrifugally excited electrostatic waves on the high energy primary beam
electrons transiting through the same region, constitutes the second step.

\item By carrying out a linear instability analysis, relevant to a
non-autonomous system, it was shown that, for physically reasonable parameters (pertinent
for example to the Crab pulsar), the characteristic timescale for the growth of Langmuir
waves is much less than the typical rotation timescales of particles. This means that the
invoked instability is extremely efficient and can rapidly convert the star slow down energy
into the electric field energy

\item Landau damping of the unstable Langmuir
waves on primary beam electrons that converts the electric field energy into particle energy is also
shown to be equally rapid. The combination creates a very efficient "machine"
that generates ultra high energy (multi TeV to PeVs) cosmic ray electrons; the star
slow down luminosity being the ultimate accelerator.
\end{enumerate}

The mechanism, developed in this paper, is rather general. The magnetospheres of active galactic nuclei
(AGNs), for example, be just as  suited to support  centrifugally driven electrostatic waves. And these waves, could, equally efficiently, accelerate any charged leptons or hadrons.

Acknowledgements: The work of SMM was supported by USDOE Contract No.DE-FG03-96ER-54366.


\begin{thebibliography}{}
%\bibitem[\protect\citeauthoryear{Abbasi et al.}{2012}]{neutr} Abbasi, R. et al. 2012,
%Nature, 484, 351
\bibitem[\protect\citeauthoryear{Abdo et
al.}{2009}]{fermel} Abdo, A.A. et al. 2009, Phys. Rev. L, 102,
181101
\bibitem[\protect\citeauthoryear{Abramowitz \& Stegun}{1965}]{abram} Abramowitz, M. \& Stegun, I.A.,
1965, {\it Handbook of Mathematical Functions,} Natl. Bur. Stand.
Appl. Math. Ser. No. 55 (U.S. GPO, Washington, D.C., 1965)
\bibitem[\protect\citeauthoryear{Ackermann et al.}{2010}]{fermel1} Ackermann, M. et al. 2010,
Phys. Rev. D, 82, 092004
\bibitem[\protect\citeauthoryear{Aharonian et at.}{2008}]{hessel} Aharonian, F., et al.
2008, Phys. Rev. L, 101, 261104
\bibitem[\protect\citeauthoryear{Arons \& Sharleman}{1979}]{arons}
Arons, J. \& Sharleman, E.T., 1979, ApJ, 231, 854
\bibitem[\protect\citeauthoryear{Blandford}{1980}]{blfermi}
Blandford, R.D., 1980, ApJ, 238, 410
\bibitem[\protect\citeauthoryear{Blandford \& Ostriker}{1980}]{blostr}
Blandford, R.D. \& Ostriker, J.P., 1980, ApJ, 237, 793
\bibitem[\protect\citeauthoryear{Blumenthal \& Gould}{1970}]{KN}
Blumenthal, G. R. \& Gould, R. J. 1970, Rev. Mod. Phys., 42, 237
\bibitem[\protect\citeauthoryear{Chang et at.}{2008}]{pamel}
Chang, J. et al., 2008, Nature, 456, 362
\bibitem[\protect\citeauthoryear{Deutsh}{1955}]{deutsh} Deutsh, A., 1955, Ann. Astrophys.
18, 1
%\bibitem[\protect\citeauthoryear{Egorenkov et al.}{1983}]{elm} Egorenkov, V.D.,
%Lominadze, D.G. \& Mamradze, P.G., 1983, Ap, 19, 426
\bibitem[\protect\citeauthoryear{Goldreich \& Julian}{1969}]{gj} Goldreich, P. \& Julian, W.H.,
1969, ApJ, 157, 869
\bibitem[\protect\citeauthoryear{Kisaka \& Kawanaka}{2012}]{shota} Kisaka, S. \& Kawanaka, N., 2012,
MNRAS, 421, 3543
\bibitem[\protect\citeauthoryear{Lominadze et al.}{1979}]{lmm}
Lominadze J.G., Machabeli G.Z. \& Mikhailovsky A.B., 1979, J. Phys.
Colloq., 40, No. C-7, 713
\bibitem[\protect\citeauthoryear{Lominadze
et al.}{1979}]{lms} Lominadze, D.G., Mikhailovskii, A.B. \& Sagdeev,
R.Z., 1979, JETP, 50, 927
\bibitem[\protect\citeauthoryear{Machabeli}{1978}]{m78} Machabeli, G.Z.,
1978, SvJPP. 4, 914
\bibitem[\protect\citeauthoryear{Machabeli et
al.}{2005}]{incr} Machabeli, G., Osmanov Z. \& Mahajan, S., 2005,
Phys. Plasmas 12, 062901
\bibitem[\protect\citeauthoryear{Machabeli \& Rogava}{1994}]{mr94} Machabeli, G.Z. \& Rogava, A.D.,
1994, Phys. Rev. A 50, 98
\bibitem[\protect\citeauthoryear{Machabeli \& Tsikarishvili}{1978}]{mt78} Machabeli, G.Z. \& Tsikarishvili, E.G.,
1978, SvJPP. 4, 920
\bibitem[\protect\citeauthoryear{Muslimon \& Tsygan}{1992}]{muslimov} Muslimon, A.G. \& Tsygan, A.I., 1992,
MNRAS, 255, 61
%\bibitem[\protect\citeauthoryear{Osmanov}{2010}]{tev} Osmanov,
%Z., 2010, New. Astron, 15, 351
%\bibitem[\protect\citeauthoryear{Osmanov}{2008}]{incr3}
%Osmanov, Z., 2008, Phys. Plasmas, 15, 032901
%\bibitem[\protect\citeauthoryear{Osmanov}{2008b}]{agnff} Osmanov, Z., 2008, A\&A,
%490, 487
%\bibitem[\protect\citeauthoryear{Osmanov et al.}{2008}]{mnras} Osmanov, Z., Dalakishvili, Z. \& Machabeli, Z.,
%2008, MNRAS, 383, 1007
\bibitem[\protect\citeauthoryear{Osmanov \& Rieger}{2009}]{or09} Osmanov, Z. \& Rieger,
F., 2009, A\&A, 502, 15
\bibitem[\protect\citeauthoryear{Osmanov et al.}{2007}]{osm7} Osmanov, Z., Rogava, A.S.
\& Bodo G., 2007, A\&A, 470, 395
\bibitem[\protect\citeauthoryear{Osmanov et al.}{2009}]{forcefree} Osmanov, Z., Shapakidze,
D. \& Machabeli, Z. 2009, A\&A, 503, 19
\bibitem[\protect\citeauthoryear{Rieger \& Mannheim}{2000}]{rm00} Rieger, F. \& Mannheim, K., 2000, A\&A, 353, 473
\bibitem[\protect\citeauthoryear{Rogava et al.}{2003}]{grg} Rogava, A., Dalakishvili, G. \&
Osmanov, Z., 2003, GReGr, 35, 1133
\bibitem[\protect\citeauthoryear{Rybicki \& Lightman}{1979}]{Lightman} Rybicki,  G.B. \& Lightman, A. P., 1979,
Radiative Processes in Astrophysics. Wiley, New York
\bibitem[\protect\citeauthoryear{Ruderman \& Sutherland}{1975}]{rud} Ruderman, M.A. \&
Sutherland, P.G., 1975, ApJ, 196, 51
\bibitem[\protect\citeauthoryear{Silin}{1960}]{silin} Silin, V.P., 1960, ZhETF,
38, 1577
\bibitem[\protect\citeauthoryear{Sturrok}{1971}]{sturrok} Sturrok,
P., 1971, ApJ, 164, 529
\bibitem[\protect\citeauthoryear{Tademaru}{1973}]{tademaru}
Tademaru, E., 1973, ApJ, 183, 625
\bibitem[\protect\citeauthoryear{Thorn et al.}{1986}]{membran}
Thorne, K., Price, R. \& MacDonald, D.A., 1986, {\it Black Holes:
The Membrane Paradigm}, (Yale University Press, New Haven, 1986)
\bibitem[\protect\citeauthoryear{Tsitovich}{1961}]{citovich} Tsitovich, V.N., 1961,
ZhETF, 40, 1775
\bibitem[\protect\citeauthoryear{Usov \& Shabad}{1985}]{usov}
Usov, V.V. \& Shabad, A., 1985, ApSS, 117, 309
\bibitem[\protect\citeauthoryear{Volokitin et al.}{1987}]{vkm}
Volokitin, A.S., Krasnoselskikh, V.V. \& Machabeli, G.Z., 1987,
Soviet Journal of Plasma Physics, 11, 310



\end{thebibliography}
\end{document}